Writing Science Fiction Stories to Motivate Analysis of Journal Articles

F. J. Kontur

For many students, college physics courses have little or even negative impact on their beliefs about the connection between physics and the everyday world.[1] One way to help students see this connection is to incorporate analysis of science articles into the course.[2-7] For introductory courses, one might have students discuss newspaper articles related to science or articles from popular science magazines while articles from peer-reviewed journals might be more appropriate for honors-level or upper-division courses. In this work, I describe a project done in a two-semester upper-division electricity and magnetism (E&M) course at the United States Air Force Academy (USAFA) in which students were required to use the science from a peer-reviewed journal article to write a science fiction story. Although others have utilized science fiction stories in physics courses,[8-13] this is the first article to describe a project in which students write their own science fiction stories in a physics course.

**Motivating Analysis of Journal Articles with a Story-Writing Exercise**

In previous semesters, I asked students in the upper-division E&M course to choose journal articles related to topics that we were covering. Groups of 3-4 students discussed these journal articles on 4 different "journal days" during the semester. After the discussion, each group presented a summary of their journal article to the rest of the class and then had to write a summary of their discussion after class. While students seemed to enjoy the journal day discussions, the actual analysis of the journal articles was, for the most part, very superficial.

To address this issue, in fall 2012, I reformulated the journal article discussion activity as a science fiction story-writing exercise. At the beginning of the semester, I asked each student to find a journal article related to a topic in the course. I told them to use the article as the basis for a short science-fiction story that they would write and turn in by the end of the semester. As with the journal discussion activity described in the previous paragraph, I set aside 4 lessons for discussion groups, but now the groups talked about journal articles in the context of the story-in-progress rather than the journal articles themselves. The days when discussions groups met created convenient checkpoints. If a student did not have the required work done for each checkpoint, then he or she would lose points. The schedule is shown in Table I.

To incentivize the story-writing activity, approximately 15% of the homework points for the course were devoted to the final story grade and successful completion of the checkpoints. A grading rubric was developed for the story and distributed to students at the beginning of the semester. The rubric described five things that would be considered in the grading of the stories (listed in order of importance):

1. It was clear from the story that the student had a good understanding of the science in the journal article.
2. The science from the journal article was an integral part of the story.

TABLE I. Schedule of activities for story-writing exercise.

| Lesson # | Work Due / Activity |
|---|---|
| 8 | Journal articles for stories must be chosen and turned in. |
| 9 | Discussion groups meet to talk about the content of each student's chosen journal article and how the journal article can be used in a story. |
| 15 | Outline of story is due |
| 16 | Discussion groups meet to talk about and critique story outlines. |
| 23 | First draft of story is due. |
| 24 | Discussion groups meet to talk about and critique 2 first drafts. Names will be removed from drafts, and the drafts will come from members of another discussion group. After the meeting, critiques will be typed up by each group and passed on to the story writer. |
| 32 | Second draft of story is due. |
| 33 | Discussion groups meet to talk about and critique 2 second drafts. Names will be removed from drafts, and the drafts will come from members of another discussion group. After the meeting, critiques will be typed up by each group and passed on to the story writer. |
| 40 | End of semester, final draft of story is due. |

3. Proper grammar and syntax made the story clear and understandable (although poor grammar may be purposely used for characterization or for narrative effect).
4. The story was well-structured and had a compelling plot.
5. The characters in the story were interesting.

    By changing the journal analysis activity in the way described, my hope was that (1) students would have more motivation for studying their article since they were creating something of their own based on their understanding of the article, (2) because science fiction is one of the more popular and accessible applications of scientific knowledge, students would have a more positive response to this activity compared to a more generic journal article analysis activity, (3) having a full semester to analyze their article would allow students to attain a high level of understanding, (4) students would have a greater appreciation of the science in their article because the activity required them to imagine possible future applications and consequences of that science. The latter point was an especially relevant one for young scientists, I believed. Funding for scientific endeavors has been coming under increasing scrutiny, and, as a result, the next generation of science practitioners must be able to make a compelling case to funding agencies and to the general public about what impacts their scientific projects will have on society.

**Student Response to Story-Writing Activity**

In the fall 2012 offering of E&M (the first semester of the two-semester course sequence), all 15 students turned in stories, and they all did reasonably well, with the class average on the final draft of the story being 84% with a standard deviation of 11%. Particularly notable was the amount of effort that students put into the stories. Even though I did not set a minimum length, all students wrote stories that were over 2,000 words. The average story length was 7,000 words, and 4 of the 15 students wrote stories that were greater than 10,000 words.

However, despite the effort that they put into writing their stories, student comments on an end-of-semester survey about the project were overwhelmingly negative. Table II gives a summary of the feedback. While more than half of the students said it was a fun and/or interesting experience, most felt that the activity was not a good use of their time and was not an appropriate activity for a physics course. Despite the negative feedback, I repeated the story-writing activity for the second semester of the course, in spring 2013. In response to the feedback from the fall semester, I placed more emphasis in the grading rubric on understanding the science of the articles and less emphasis on writing a good story. Also, I had a short talk with students before each of the discussion days about what I wanted them to get out of the story-writing activity. I hoped that this would reinforce the purpose of the assignment. Nevertheless, the feedback at the end of spring 2013 was substantially the same as in fall 2012, and 2 of the students did not even bother turning in a final draft of the story.

Many of the students said that they believed their learning was hindered by having to spend time doing the creative writing assignment. While it is impossible to prove definitively whether or not this was the case, I did administer a standardized exam, the Colorado Upper-Division Electrostatics diagnostic,[14] to students at the beginning and end of the fall 2012 semester. On the pre-test, they scored 24%, which is 6-19% below pre-test scores for students in equivalent courses.[14] The likely reason for this is that, because of the schedule of courses at USAFA, it is often more than a year between upper-division E&M and any previous courses that physics majors have had on E&M topics. Nevertheless, despite their low pre-test scores, they scored 51% on the post-test, which is equal to the average post-test score for upper-division E&M students at other institutions.[14] Therefore, there does not seem to be any evidence that student learning was adversely affected by the story-writing activity compared to students at other schools taking the same type of course without such an activity.

**Lessons Learned**

*Lesson 1: Make sure that the activity you have chosen is connected in a substantial way to course learning objectives and make sure that there are quiz and exam questions related to the activity. Emphasize these connections to students.*

There are interesting parallels between the issues encountered in the story-writing activity and the issues that have been encountered when using blogs in science courses.[15-20] Research has found that if students view the blog assignments as a side-activity that is not related to the rest of the course, then they will choose not to participate or will participate only in a very superficial way.[15,16,19] The obvious reason for this is that students believe the activity is irrelevant to their performance and learning in the rest of the course. Several of the comments in Table II reflect such a belief about the story-writing activity, describing it as something that "was a waste of time", "isn't relevant to the [exams]", and "takes too much time away from the course objectives."

TABLE II.   Student feedback on story-writing activity.

| Question | Representative Examples of Responses |
|---|---|
| Looking back at your story writing experience, what were your initial reactions to the assignment? | - I really don't like writing, that's why I'm not good in English. It kind of seemed out of place for a physics class.<br>- Really? Why? If I had the time, sure, but I can't even really keep up with homework.<br>- I thought it would be fun but time consuming.<br>- Initially, I dreaded it a little, but only because I'm not a big writer, and definitely not a creative writer. Reading the journal article wasn't the part I dreaded. |
| Did the research that you did for your creative writing assignment make you more interested in the topics covered in E&M? Why or why not? | - Yes. [E&M] is helping me understand (somewhat) actual research going on at the cutting-edge in physics.<br>- Can't say it did. Just another long obligation for an assignment.<br>- Not particularly. Would have been better if we worked through the article as a class or had an experiment/demo. |
| What were your most common feelings about the project as you completed the various components? | - I thought it was fun until I had to write the beast.<br>- Doing this is a lot of work which isn't relevant to the [exams].<br>- Writing was labersome. I am a physics major because I dislike subjective and open ended papers!<br>- They were fun because writing a story is interesting, but it was still a waste of time. |
| Now that you have finished writing your story, how do you feel about the experience? Do you have any suggestions for improving this project for future semesters? | - Honestly, I'd say scrap it. There are so many concepts I'm still uneasy about in the class that I much rather would've learned instead.<br>- It was fun and creative writing the story, but I find it weird that my longest paper at the Academy was for Physics.<br>- I think the project should be disregarded because it takes too much time away from the course objectives.<br>- It was a cool assignment. My only suggestion would be to be more lenient on what articles we can use. |

*Lesson 2: If there are several different assignments that can achieve the same learning goals just as effectively, allow students to choose which assignment they want to do unless there is a good reason not to do so.*

As described in the previous section, my intention with the story-writing activity was to create a fun learning experience that would motivate students to make a good effort to understand a journal article.

Unfortunately, I did not considered what would happen if students did not find the activity fun or motivational. Several students pointed out in their feedback that they understood that having them analyze a journal article was the goal of the activity, but that they disliked the creative writing aspect and would have preferred to do an assignment that involved a straightforward analysis of a journal article. Certainly a lot of the negative feedback on the story-writing activity could be traced back to student resentment at being forced to do an assignment that was outside of their comfort zone.

*Lesson 3: For an open-ended assignment, try to give as much explicit guidance as possible on expectations about time spent on the assignment. Specifically, try to be aware of students who may try to go overboard and counsel them about more efficient ways to meet the goals of the assignment while not discouraging their enthusiasm.*

Many students commented on what they considered to be the excessive time commitment of the story-writing assignment. While this is not a bad thing, per se, it is clear from their feedback that at least a few students felt that they spent a majority of their time trying to write a good story rather than learning about science. For example, students who wrote 10,000+ words would probably have been better-served by being forced to rein in their efforts.

*Lesson 4: Be aware of student expectations about what should be and what should not be in a physics course and make a plan about how to handle those expectations.*

Many of my students believed, perhaps with some justification, that a creative writing assignment was inappropriate for a physics course. More appropriate learning activities, according to them, would have been homework, exams, experiments, and demonstrations. This is a not a new phenomenon. As Redish *et al.* say, "We are frustrated by the tendency many students have to… spend a large amount of time memorizing long lists of uninterpreted facts or performing algorithmic solutions to large numbers of problems without giving any thought or trying to make sense of them."[21] In other words, while teachers may feel it is important for students to learn that physics is more than just a collection of equations, many students do not see the importance or the purpose of such learning. The obstacles presented by such student expectations are not impossible to overcome, but non-traditional course activities will have a greater likelihood of success if teachers are aware of these expectations beforehand and have a plan for how to deal with them.

**Conclusion**

Based on student feedback, the first iteration of the story-writing activity was not successful in motivating students to read scientific articles and apply that knowledge to what they were learning in the course. However, their feedback suggests that such an activity can be made successful by incorporating it into other aspects of the course, particularly assessments. Also, students suggested in their feedback that they would have liked more options for how to achieve the learning objectives of the activity. Specifically, some students might have been more motivated by a traditional article analysis activity while other students might have been more enthusiastic about a creative activity. In general, one would expect more student buy-in for an activity that they have chosen rather than one that is forced upon them. Finally, I hypothesize that a version of the story-writing activity, perhaps involving scientific articles from popular news sources, could be even more successful in an introductory physics course than an upper-division course. This is because students who are not physics majors will likely not feel as much resentment about taking time out of "regular" physics class activities. Indeed, introductory students will likely enjoy the change of pace from day-to-day classwork.

**Notes**

Email me at frederick.kontur@usafa.edu for examples of stories written by students for this activity.

Distribution A, approved for public release, distribution unlimited.